\documentclass[9pt,twocolumn,twoside]{pnas-new}

\usepackage{soul,xcolor}

\usepackage{siunitx} 
\DeclareSIUnit\Molar{\textsc{M}}
\usepackage{verbatim} 
\usepackage[caption=false]{subfig}

\templatetype{pnasresearcharticle} 

\begin{document}
\setstcolor{red}

\title{A microfluidic band-pass filter for flexible fiber separation}

\author[a,1]{Zhibo Li}
\author[a,b,1]{Clément Bielinski}
\author[a,c,2]{Anke Lindner}
\author[b,2]{Blaise Delmotte}
\author[a,2]{Olivia du Roure}

\affil[a]{Laboratoire de Physique et Mécanique des Milieux Hétérogènes, École Supérieure de Physique et de Chimie Industrielles de la ville de Paris,
Université Paris Sciences et Lettres, Sorbonne Université, Université Paris Cité, CNRS, 75005 Paris, France}
\affil[b]{LadHyX, CNRS, École Polytechnique, Institut Polytechnique de Paris, 91120 Palaiseau, France}
\affil[c]{Institut universitaire de France (IUF), France}

\leadauthor{Li}

\significancestatement{
Flexible fibers are common in nature and technology, from biological filaments to nanomaterials and pollutants. However, sorting these deformable, anisotropic objects in flow remains difficult. We show that microfluidic devices with tilted pillar arrays—originally designed for spherical particles—can be adapted to separate fibers by length. Unexpectedly, these arrays act as band-pass filters, causing lateral migration only for fibers whose lengths match the inter-pillar spacing. This selectivity arises from a complex interplay between fiber deformation, tension from the flow, and interactions with obstacles. Our findings reveal a novel physical mechanism for controlling the transport of soft, elongated objects and open new possibilities for precision sorting in fields like biotechnology, environmental science, and materials engineering. 
}

\authorcontributions{Z.L., C.B., A.L., B.D. and O.d.R. designed research, Z.L. performed experiments, C.B. performed simulations, Z.L. and C.B. performed data curation, Z.L., C.B., A.L., B.D. and O.d.R. analyzed data, Z.L. and C.B. wrote the original draft, Z.L., C.B., A.L., B.D. and O.d.R. reviewed and edited the final version.}
\authordeclaration{The authors declare no competing interests.}
\equalauthors{\textsuperscript{1}Z.L contributed equally to this work with C.B.}
\correspondingauthor{\textsuperscript{2}To whom correspondence should be addressed. E-mail: anke.lindner@espci.fr, blaise.delmotte@cnrs.fr, olivia.duroure@espci.fr}


\begin{abstract}

The control of particle trajectories in structured microfluidic environments has significantly advanced sorting technologies, most notably through deterministic lateral displacement (DLD). While previous works have largely targeted rigid, near-spherical particles, the sorting of flexible, anisotropic objects such as fibers remains largely unexplored. Here, we combine experiments and simulations to demonstrate how tilted pillar arrays enable efficient, length-based separation of flexible fibers. 
We discover that these arrays act as band-pass filters, selectively inducing lateral migration in fibers whose lengths are close to the array period. Fibers significantly shorter or longer exhibit small lateral deviation. This migration arises from the interplay of fluid-structure interactions between fibers and the complex flow and steric interactions with the pillars. Depending on their length, fibers exhibit distinct transport regimes: short fibers zigzag in between pillars following the flow, intermediate length fibers exhibit wrapping and jumping from one pillar to another, leading to lateral displacement, and long fibers deform extensively, following mixed zigzag-jump trajectories with small lateral migration. 
We identify the mechanical tension that develops in the fiber when wrapped around the pillars as the driving mechanism of cross-streamline transport. Leveraging this band-pass effect, we designed a highly efficient separation device to collect monodisperse fiber suspensions. Our findings not only expand the functional scope of DLD-like systems but also open new avenues for understanding transport of anisotropic objects in porous media.

\end{abstract}

\dates{This manuscript was compiled on \today}
\doi{\url{www.pnas.org/cgi/doi/10.1073/pnas.XXXXXXXXXX}}

\maketitle
\thispagestyle{firststyle}
\ifthenelse{\boolean{shortarticle}}{\ifthenelse{\boolean{singlecolumn}}{\abscontentformatted}{\abscontent}}{}

\firstpage[4]{5}


\dropcap{T}he ability to sort and separate particles is of fundamental importance. It plays a crucial role in various fields, such as medical diagnostics, where blood is separated for specific clinical purposes \cite{ward2011conventional, loutherback2012deterministic, mielczarek2016microfluidic}; biological analysis, such as in chromatography before DNA sequencing \cite{huang2002dna, dorfman2010dna}; and environmental monitoring, like in water 
or air
filtration processes \cite{padervand2020removal,liu2021review,de2023fibrous}.

The rapid advancement of microfluidics has introduced innovative particle separation techniques, including pinched flow fractionation, microvortex manipulation, and various filtration methods \cite{sajeesh2014particle}. Among these, deterministic lateral displacement (DLD) stands out for its label-free, continuous separation capabilities and high resolution \cite{McGrath2014, Salafi_2019, Hochstetter_2020}. 
Since the seminal work of Huang \textit{et al.} in 2004 \cite{Huang2004}, DLD has attracted significant attention for its effectiveness in separating near-spherical particles by size, as well as small deformable particles, such as red blood cells, by size or deformability  \cite{Huang2004,Loutherback2010,Beech2012,Zhang2015,Henry2016}. The principle is straightforward: particles interact with a periodic pillar array, and depending on their size relative to a critical threshold, they follow different streamlines and are either laterally displaced or carried along the main flow direction. The critical size for separation is determined by the pillar geometry and the incidence angle of the array relative to the flow direction  $\alpha$ \cite{Davis2008MicrofluidicDisplacement}.

\par
Studies on flexible polymers such as DNA in DLD devices are extensive, and due to the coiled state they adopt at rest, they have mostly been treated as soft spheres  \cite{Huang2004, Chen2015, strom2022high}. Their dynamics under flow (such as stretching and rotation) are treated as secondary effects \cite{strom2022high}. The use of DLD for elongated and deformable particles, such as microplastics, textile fibers, or elongated parasites, remains however underexplored, despite its benefits for environmental assessments, decontamination and biomedical applications.   
\par
The work most closely related to flexible fibers in periodic micropillar arrays is the 2D numerical study of Chakrabarti \textit{et al.},  which  demonstrated the potential for the chromatographic sorting  of Brownian fibers by length  \cite{chakrabarti2020trapping}.  Yet, in this work, continuous, high-throughput, separation  using DLD has not been explored and experimental validation of this application is lacking. 

The dynamics of  flexible fibers in viscous flows are primarily governed by the elastoviscous number, $\bar{\mu} = 8\pi\mu\dot{\gamma}L^4/B c$, which quantifies the balance between the viscous forces exerted by the flow and the internal elastic restoring forces of the fiber \cite{du_roure_dynamics_2019}. Here, $\mu$ is the solvent viscosity, $\dot{\gamma}$ is the shear rate in simple shear flow, $L$ is fiber contour length, $B$ is bending rigidity, and $c = -\ln(\epsilon^2 e)$ is a dimensionless slenderness parameter, with $\epsilon$ representing the fiber aspect ratio, and $e$ Euler's number.  In our study the Reynolds number, ${\rm Re} = \rho U_{\rm 0}L /\mu$, is of the order of $10^{-4} \ll 1$ and inertia is negligible. 
For spatially uniform $\dot{\gamma}$, the dynamics of a flexible fiber is well-understood: the critical elastoviscous number for buckling instability is $\bar{\mu} = 306.4$ in simple shear flows and $\bar{\mu} = 153.2$ in straining flows \cite{Quennouz2015,du_roure_dynamics_2019}. However, within a pillar array, the presence of obstacles generates a more complex flow field with curved streamlines, combining various flow types and causing spatial variations in $\dot{\gamma}$. In such conditions, fiber behavior becomes more involved, as they may experience buckling instabilities or simple bending in response to the complex flow and fiber obstacle interactions, making it challenging to fully understand their dynamics and the associated separation potential.

In this study, we perform experiments using actin filaments transported in microfluidic channels. Actin filaments results from the self-assembly of monomeric protein actin into filaments which are typically tens of micrometer long. They can be used as model systems for flexible Brownian filaments because the persistence length is close to their typical lengths \cite{Liu2018MorphologicalFlow,chakrabarti2020flexible}. Fluorescently labeled, their conformation and transport can be observed in viscous micro-flows.  The experiments are combined with numerical simulations to investigate the separation of flexible fibers using a square  periodic lattice of cylindrical pillars  under varying tilt angles $\alpha$ (see Fig.~\ref{fig:setup}\textit{A}). 
In the simulations, immersed fibers are modeled with a bead-spring model and coupled to the ambient flow, which is computed with the lattice Boltzmann method (LBM), with an immersed boundary kernel and a variant of the Stokesian Dynamics method \cite{Wajnryb2013}. Our model accounts for  the fibers feedback on the flow through the hydrodynamic interactions (HI) between the fiber
beads; but it neglects the corrections of these HI due to the channel walls and obstacles’ surface, which induce small velocities compared to the fiber velocity induced by the ambient flow \cite{Li2024} and to the lubrication corrections in the near field. Brownian motion is ignored and near-field interactions with obstacles are accounted for using lubrication forces and a repulsive potential (see Methods for more details).
In our experiments, all fibers have a length $L > 5$ \unit{\um}, which,  using a shear rate $\dot{\gamma} = U_{\text{mean}}/(\lambda - 2R)$ based on the average fluid velocity in the channel and the gap between pillars, 
results in an elastoviscous number $\bar{\mu}$ that can exceed $10^3$. This value is  well above the instability threshold, and fibers can therefore be strongly deformed by the flow.

Our combination of experiments and simulations in the pillar arrays reveals that flexible filaments can exhibit a wide variety of migration patterns controlled by their length relative to the lattice spacing, $L/\lambda$, where $\lambda$ is the size of the unit cell of the periodic lattice. In particular, our  results show that, at a tilt angle $\alpha = 35^\circ$, repeated interactions with the obstacles induce periodic deformations and large lateral migrations of the fibers across the microfluidic chip but only within a certain length range close to $L/\lambda\sim1$. This lateral migration at intermediate lengths occurs because fibers wrap around pillars and, as their free end is pulled by the flow, migrate across streamlines due to the buildup of internal tension and prevents them from aligning with the flow. Outside this length range, fibers are mostly transported with the flow and exhibit weak lateral migration. Short fibers  barely interact with obstacles, while long fibers are intermittently strongly deformed by their interactions with different pillars simultaneously, which intermittently reduces the apparent length.  

By triggering these length-dependent dynamics, the pillar array acts as an accurate  ``band-pass filter'' for sorting flexible fibers by length in microfluidic chips. The very good agreement between experiments on Brownian actin filaments and numerical simulations of non-Brownian flexible fibers supports the universality of this finding, which holds true for any anisotropic structure, Brownian or non-Brownian, regardless of its specific details.
This band-pass effect, which is not observed in conventional DLD for spherical particles, opens promising perspectives for particle separation and filtering.

\section*{Results}
\subsection*{Typical fiber dynamics in the array}
\label{sec:fiber_transport}

\begin{figure*}
    \centering
    \includegraphics[width=17.8cm]{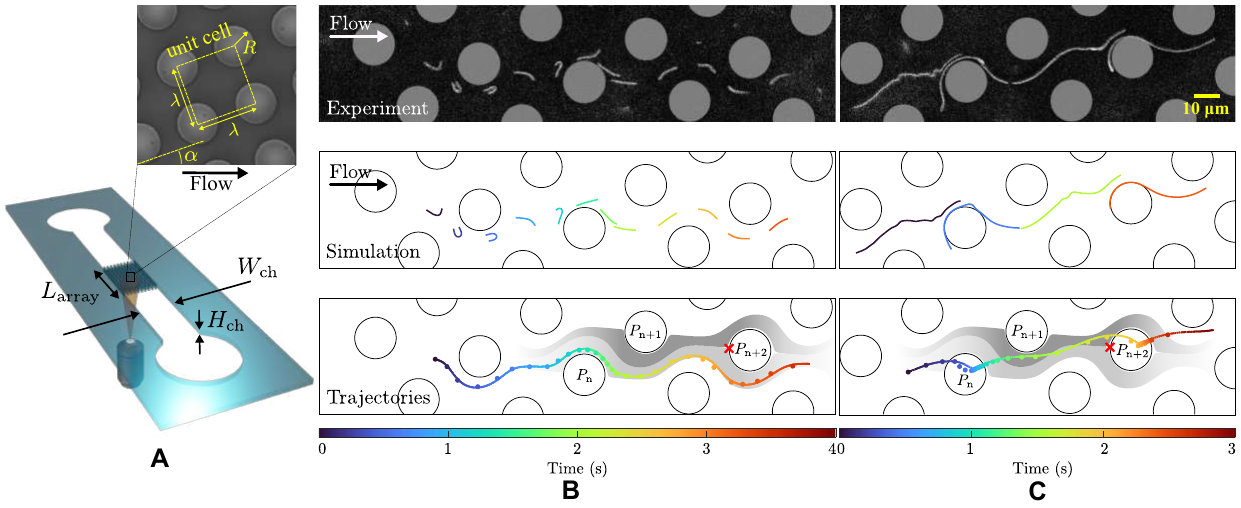}
    \caption{Geometry of the channel and examples of fiber dynamics. (\textit{A}) A 3D schematic of the experimental channel (not to scale), with a magnified top-view bright-field image of the pillar array shown alongside. In the experiments, $H_{\rm{ch}}= 50\,\unit{\um}$, $R=10\pm1\,\unit{\um}$ and $\lambda=30\,\unit{\um}$ (these  geometric parameters are identical in numerical simulations, see Methods). 
    (\textit{B} and \textit{C}) Chronophotographs of fibers from experiments (top row) and simulations (middle row) at a tilt angle between flow and array directions of $\alpha = 35^\circ$. The bottom row presents the corresponding fiber CoM trajectories extracted from the simulation (solid lines) and experiment (dots). Colors indicate the time. The shaded gray regions indicate the flow lanes between pillars $P_{\rm{n}}$ and $P_{\rm{n+1}}$, separated by a stagnation point at pillar $P_{\rm{n+2}}$ (red cross).  (\textit{B}) The fiber length is $L = 9.6 \,\unit{\um} = 0.32\lambda$. (\textit{C}) The fiber length is $L = 48 \,\unit{\um} = 1.6\lambda$. In both cases, the initial conditions for the simulations are extracted from experiments. In the experiments actin filaments are suspended in an aqueous solution of  $45.5\,\unit{\percent}$ (w/v) sucrose, with a viscosity of $\mu = 5.6\,\unit{\milli\pascal\second}$. Note that the fiber CoM can be located inside the pillar when it is wrapped around. }
    \label{fig:setup}
\end{figure*}

When flowing in a periodic square array, flexible fibers exhibit complex conformations and migration patterns that depend on their length. Fig.~\ref{fig:setup}\textit{B} and \textit{C} show two examples of such complex dynamics at a tilt angle $\alpha = 35^\circ$ with two representative fiber chronophotographs from experiments (top row) and simulations (middle row). In Fig.~\ref{fig:setup}\textit{B}, a fiber with a contour length of $L= 9.6$\,\unit{\um} $ = 0.32\lambda $ undergoes significant deformation and tumbling in the regions of high shear rate near the pillars but passes freely between the pillars without making contact.  Initially adopting a U-shape just below the first obstacle, it subsequently unfolds and exhibits strong deformations further downstream. 
In contrast, a longer fiber with $L = 48$\,\unit{\um} $= 1.6\lambda$ wraps around the pillars and stays temporarily trapped in their vicinity before being released into the surrounding flow to continue towards successive pillars. It is worth noting that parts of the filament  experiencing high compressive forces from the flow can buckle as illustrated by the first shape of Fig.~\ref{fig:setup}\textit{C}. 
The excellent agreement between experiments and simulations for these two examples is easily visible in terms of morphology on the two sets of chronophotographs (first two rows of Fig.~\ref{fig:setup}\textit{B} and \textit{C}). In addition, the experimental and simulated trajectories of the fiber center of mass (CoM) superimpose almost perfectly (bottom row of Fig.~\ref{fig:setup}\textit{B} and \textit{C}, where time is color-coded) emphasizing that the simulations capture the details of the fiber dynamics.

The comparison of the fiber trajectories (bottom row of Fig.~\ref{fig:setup}\textit{B} and \textit{C}) already reveals distinct behaviors, which could be leveraged for separation. The shorter fiber primarily moves in the flow direction, meandering between pillars. In contrast, the longer fiber wraps around the pillars and exhibits noticeable lateral migration even within this small section of the array. To facilitate the discussion, we have labeled three pillars, $P_{\rm{n}}$,  $P_{\rm{n+1}}$ and $P_{\rm{n+2}}$. The pillar $P_{\rm{n+2}}$ marks the point where the trajectories of the two fibers deviate, the longer one passes above $P_{\rm{n+2}}$ while the shorter one is transported below. We identify two groups of streamlines which separate at the stagnation point of $P_{\rm{n+2}}$ (denoted by a red cross) and which are marked in different shades of gray at the bottom of Fig.~\ref{fig:setup}\textit{B} and \textit{C}. 
The trajectory of the shorter fiber is confined within one of these regions (light gray) and the fiber is transported along a given streamline as a tracer would be. 
In contrast, the longer fiber, initially located in the light gray region finally exhibits streamline crossing and reaches the dark gray region in the vicinity of pillar $P_{\rm{n+2}}$. In the next section we will describe the flow in the pillar array as a first step in the understanding of fiber dynamics. 

\subsection*{Flow in the pillar array}
\label{sec:Flow}

Despite the regular arrangement of pillars, the flow in a tilted array is complex and features regions dominated by simple shear, elongational, and rotational components \cite{haward2021stagnation}.
 Since the flow is in the Stokes regime, the periodicity of the array is reproduced in the flow pattern, which is completely described by the flow within a unit cell. In Fig.~\ref{fig:flow_poincare_map}\textit{A}, we compare flow fields in the unit cell measured using micro-particle image velocimetry (\textmu{PIV}) and simulated using LBM \cite{Krueger2016,Li2024} for different tilt angles, $\alpha$, between the flow and the array orientations, increasing from $\alpha=0^\circ$ to $\alpha=45^\circ$. The velocity magnitudes and streamlines show excellent agreement between experiments and numerical simulations.

For an array aligned with the flow direction ($\alpha=0^\circ$), the streamlines enter the unit cell from one side and exit from the opposite side. If the array is tilted by $\alpha=45^\circ$, streamlines enter by two different sides of the unit cell, the left and top sides, and exit symmetrically by the two other sides. 
For intermediate values of $\alpha$, the streamlines entering from the left side split into two groups. One group (shown in red) exits through the bottom side, as seen for $\alpha = 45^\circ$. The other group (in blue) merges with streamlines coming from above and exits through the right side.

\subsection*{Poincaré maps to assess fiber separation potential}
\label{sec:Poincare}

\begin{figure*}
    \centering
    \includegraphics[width=15cm]{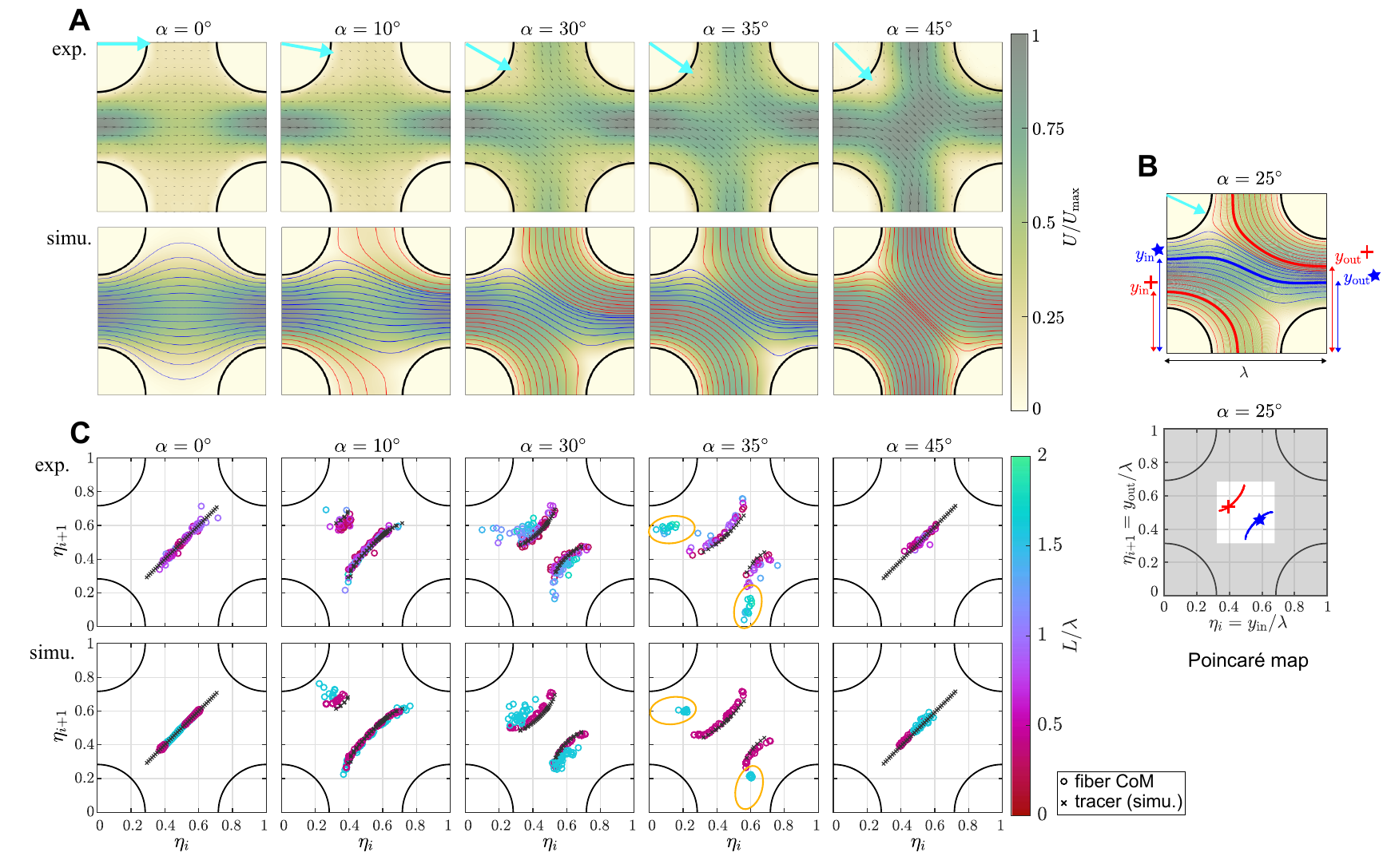}
    \caption{Characterization of the flow field and Poincaré map in a unit cell. 
    (\textit{A}) Velocity field at various tilt angles $\alpha$ measured by µPIV (top) and simulated by LBM (bottom). Colors indicate the velocity magnitude normalized by the maximum value in each unit cell, while arrows and curves represent the streamlines. The cyan arrows are the flow directions 
    (\textit{B}) Definition of Poincaré maps. The gray region indicates the non-accessible area on the Poincaré map of the tracers. The normalized position (scaled by the unit cell size $\lambda$) of a streamline as it enters the unit cell is ($\eta_{\rm i} = y_{\rm{in}}/\lambda$)  and  ($\eta_{\rm i+1} = y_{\rm{out}}/\lambda$) as it exits the unit cell  \cite{Kim2017}.
    (\textit{C}) Poincaré maps of the fiber CoM in the experiments (top row) and simulations (bottom row) at various tilt angles $\alpha$. Crosses show the Poincaré maps of streamlines computed from LBM simulations. Colors correspond to the contour length of the fibers, with two lengths shown in the simulation plots: $L=0.4\lambda$ and $L=1.6\lambda$. 
    Orange ellipses at $\alpha = 35^\circ$ outline the separation of longer fibers from shorter ones and from streamlines. Quarter circles in each corner of the plots represent the surface of the pillars.
    The Poincaré maps from experiments were derived from over 270 fiber trajectories, each comprising more than 50 fiber instances reconstructed from image frames.
    }
    \label{fig:flow_poincare_map}
\end{figure*}

To assess the separation potential of our pillar array and to study the impact of the tilt angle $\alpha$, we observed many filaments transported in a fixed portion of the array, typically formed by four to five rows of pillars in the experiments (see Fig.~\ref{fig:setup}\textit{B} or \textit{C}, top row). Using Poincaré maps \cite{Kim2017}, the filament CoM position at the entry to a unit cell is directly associated with the position at the exit, which is the entry position of the next cell (Fig.~\ref{fig:flow_poincare_map}\textit{B}). We can then compare the behavior of filaments and tracer particles independently of the details of their trajectories within the cell to evaluate the separation potential at a given tilt angle $\alpha$. Fig.~\ref{fig:flow_poincare_map}\textit{B} illustrates how the Poincaré map is built for a tracer particle plotting $\eta_{\rm i+1} = y_{\rm{out}}/\lambda$  as a function of $\eta_{\rm i} = y_{\rm{in}}/\lambda$.  This mapping is specific to the flow field of a given pillar array.

The Poincaré maps of the tracers are represented by cross markers in Fig.~\ref{fig:flow_poincare_map}\textit{C}, and capture well the grouping of the streamlines that we discussed earlier: when the tilt angle is  $\alpha=0^\circ$ or $\alpha=45^\circ$, all points lie along the line $\eta_{\rm i} = \eta_{\rm i+1}$, indicating that the inlet and outlet positions are identical for all streamlines. 
The maps split into two distinct branches for the other values of the tilt angles, which are separated by the line $\eta_{\rm i} = \eta_{\rm i+1}$. 

Fig.~\ref{fig:flow_poincare_map}\textit{C} also shows the Poincaré maps of the fiber CoM for experiments (top row) and simulations (bottom row), with a color-code for the fiber length. 
For $\alpha = 0^\circ$ and $\alpha = 45^\circ$, all data points align along the line $\eta_{\rm i} = \eta_{\rm i+1}$, regardless of fiber length,  indicating that all fibers are transported along the streamlines similarly to tracer particles. 
On the contrary, at $\alpha=30^\circ$ and $\alpha=35^\circ$, fibers with lengths on the order of $\lambda$ separate from the others (as indicated by the orange ellipses for $\alpha=35^\circ$ in Fig.~\ref{fig:flow_poincare_map}\textit{C}). This demonstrates length-dependent trajectories which can be leveraged for fiber suspension separation.

As the fiber migration takes place downstream of the wrapped pillar, the widths of the flow lanes (shown in dark and light gray in Fig.~\ref{fig:setup}\textit{B} and \textit{C}) play a crucial role in fiber separation. Simple geometric arguments based on mass conservation indicate that a tilt angle of $\alpha = 35^\circ$ offers optimal conditions, yielding comparable widths for the two flow lanes two columns downstream from the wrapped pillar (see \textit{SI Appendix}, Fig. S2). This configuration ensures that the light gray  lane is sufficiently wide to confine shorter fibers, while the dark gray lane provides adequate space for longer fibers to be collected. We will consider this pillar array orientation in the following. 
 Before exploring the potential for  fiber suspension separation at $\alpha = 35^{\circ}$, we will discuss the lateral migration mechanism in the next section.

\subsection*{Filament tension is the key for cross-streamline migration}
\label{sec:tension}

\begin{figure*}
    \centering
    \includegraphics[width=15cm]{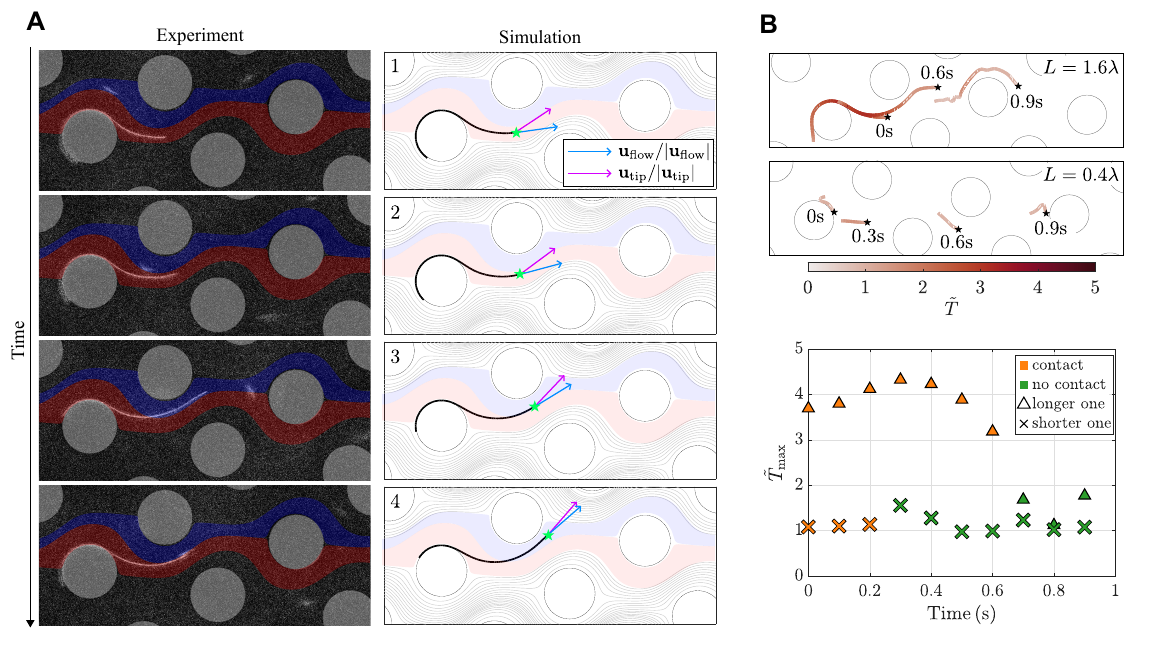}
    \caption{Cross-streamline migration during the wrapping phase of the fiber around a pillar at a lattice tilt angle $\alpha=35^\circ$. (\textit{A}) Experimental and numerical snapshots captured during the wrapping phase. In the numerical snapshots, arrows indicate the velocity directions of both the fiber tip (marked with a star) and the flow at the tip's position. 
    (\textit{B}) The upper panel displays chronophotographs of a longer and a shorter fiber, with normalized tension represented by a color scale. Black stars indicate the downstream leading ends of the fibers.
    Notice that the CoM of the longer fiber moves more slowly than that of the shorter one.
    The lower panel presents the maximum normalized tension over time corresponding to the chronophotographs above. Orange indicates periods when the fibers are in contact with the pillars, while green indicates no contact.  
    } 
    \label{fig:tension_during_wrapping_phase}
\end{figure*}

To gain a better understanding of the separation mechanisms, we examine the dynamics of a fiber that has wrapped around a pillar and became temporarily trapped. In Fig.~\ref{fig:tension_during_wrapping_phase}\textit{A}
we show sequential shapes as the flow pulls on the free end and slowly detaches the fiber which slides along the pillar. Two groups of streamlines, separated by the stagnation point on the pillar located two columns away from the wrapped pillar, are outlined in blue and red in Fig.~\ref{fig:tension_during_wrapping_phase}\textit{A}. They are similar to those indicated by a shade of gray in Fig.~\ref{fig:setup}\textit{B} and \textit{C}. It is evident that between the third and fourth snapshot the free end migrates from the red group of streamlines to the blue group (Fig.~\ref{fig:tension_during_wrapping_phase}\textit{A}). This cross-streamline migration is also visible from the simulation results in the middle column of the figure, where the velocity direction of the fiber end (last bead in the simulations, magenta arrow) is clearly not aligned with the local flow direction (cyan arrow).

Flexible elastic filaments anchored at one end in a viscous flow have been shown by Autrusson \textit{et al.} \cite{autrusson2011shape} to deviate from perfect alignment with curved streamlines. This behavior arises from internal tension developing in the inextensible filament as viscous forces pull on it, preventing it from conforming to the curvature of the streamlines and effectively stiffening the filament. 
In the simulations, we can measure the normalized tension $\tilde{T}$ by computing the stretching forces acting on each spring, normalized by the viscous drag force acting on one bead. At rest this tension is found to be close to one as steric repulsion forces present in the simulations to avoid overlap of the beads have been chosen to scale with the same viscous drag force (See Methods).  In the upper panel of Fig.~\ref{fig:tension_during_wrapping_phase}\textit{B}, we present chronophotographs of a shorter and a longer fiber, with tension indicated by color-coding. This reveals that the longer fiber experiences higher local tension than the shorter fiber, especially when wrapped around the pillar.
This is also clearly illustrated in the lower panel of Fig.~\ref{fig:tension_during_wrapping_phase}\textit{B}, which presents the maximum normalized tension along the fiber over time. The shorter fiber maintains low tension all along its transport regardless of contact (crosses; orange: contact, green: no contact). In contrast, the longer fiber experiences high values of tension during wrapping (orange triangles), significantly larger than those experienced in the absence of contact (green triangles). As discussed earlier, tension prevents flexible filaments to conform to curved streamlines and allows the fiber tip to deviate from the local flow direction, thereby facilitating lateral migration of the fiber. It is interesting to note that here tension effectively ``rigidifies" the fibers and plays a role analogous to steric hindrance in classical DLD methods for spherical objects.


\subsection*{A pillar array tilted by $\alpha = 35^\circ$ acts as a band-pass filter}
\label{sec:bandpass_effect}
The prior sections discussed the potential for fiber separation at $\alpha = 35^\circ$ based on observations on the scale of a unit cell. 
To assess the separation efficiency at the scale of a full microfluidic chip for fiber suspensions, we conduct chip-scale experiments (see \textit{SI Appendix}, Fig. S1) and simulations as explained in the following.
In the experiments, we designed a specific chip with a flow-focusing device to inject the fiber suspensions into the pillar array as a narrow band near one of the channel walls so that the deviation of migrating fibers is oriented toward the other wall. We are thus able to determine fiber length distributions at a fixed downstream location $x_{\rm f}$ but in different lateral positions $y_{\rm f}$. We  define the migration angle $\beta=\arctan(y_{\rm f}/x_{\rm f})$ and we plot its value as a function of the binned fiber length to confirm the separation ability of our device (see Fig.~\ref{fig:separation_vs_length}A). We also performed experiments in a different $\alpha = 35^\circ$-tilted array (orange symbols) in which $\lambda$ and $R$ were increased keeping the ratio $\lambda/R$ constant ($\lambda$ was increased from $30$ to $45 \mu$m, $R$ from $10$ to $15\mu$m). The relationship between the migration angle $\beta$ and the normalized fiber length $L/\lambda$ follows the same trend, even though data for very long fibers is limited by the maximum achievable length of actin filaments.

For the numerical counterpart, we perform long-time simulations using periodic boundary conditions (see Methods). 
For each value of the fiber length $L$, 50 different runs are initialized with different initial fiber configurations extracted from the experiments. The lateral position $y_{\rm f}$ of the fiber CoM after letting the fiber evolve until $x_{\rm f} \approx 300 \lambda$, comparable to the experiments 
is then averaged over the ensemble of the 50 runs and used to compute the corresponding value of $\beta$ (see Fig.~\ref{fig:separation_vs_length}B).
Since the experiments involve a relatively dilute suspension ($50\,\unit{\nano\Molar}$, approximately $0.002\,\unit{\milli\gram\per\milli\liter}$), which is well below the semi-dilute transition concentration of $0.05\,\unit{\milli\gram\per\milli\liter}$ for actin filament suspension \cite{jammey1994mechanical}, the simulations model a single isolated fiber per run, neglecting potential hydrodynamic interactions between fibers.


Fig.~\ref{fig:separation_vs_length} reveals a non-monotonic relationship between the migration angle, $\beta$, and fiber length, $L/\lambda$, for both experiments and simulations, with the emergence of three distinct regimes:

\begin{figure*}
    \centering
    \includegraphics[width=\textwidth]{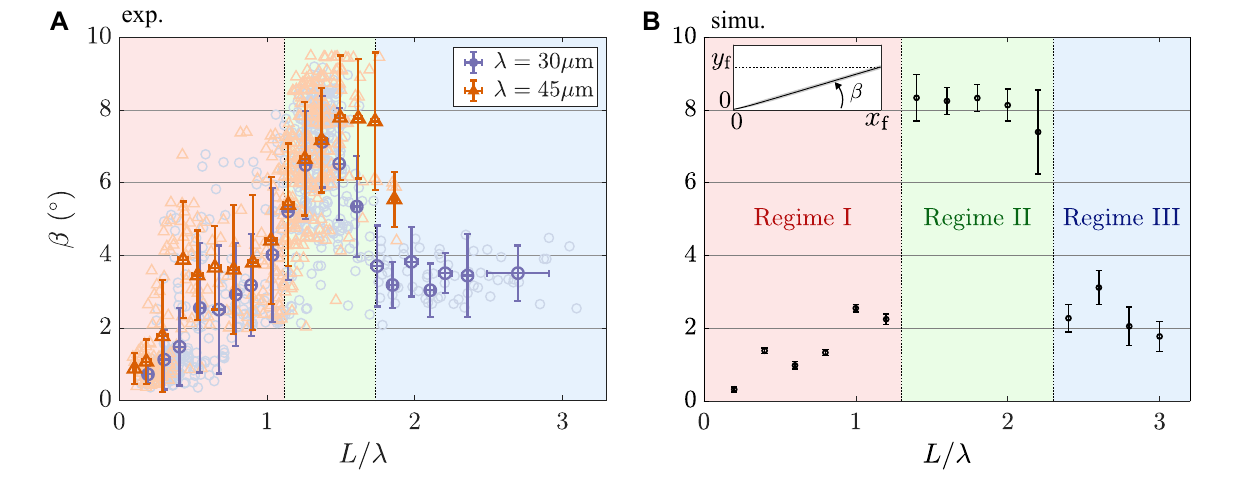}  
    \caption{
    (\textit{A}) Experimental relationship between fiber lateral migration angles $\beta$ and fiber lengths from experiments using the separation setup shown in \textit{SI Appendix}, Fig. S1, with $\lambda/R=3$. The error bars denote the group average with a bin width of $\Delta L/\lambda = 0.12$.
    (\textit{B}) Simulated angles of migration $\beta$ of the fibers as a function of their length. Data are averaged over 50 runs for each length. Vertical black lines represent the standard deviation. Inset shows the definition of the migration angle $\beta=\arctan\left(y_{\rm f}/x_{\rm f}\right)$, with $y_{\rm f}$ the lateral position of the fiber CoM at $x_{\rm f}$ downstream from the start of the pillar array. Black line is the averaged fiber CoM trajectories over the 50 runs. 
    }
    \label{fig:separation_vs_length}
\end{figure*}

$\bullet$~In the first regime, referred to as regime I, where $L < 1.3\lambda$ according to simulations and experiments, the migration angles of the fibers show a moderate increase with increasing length. However, they remain small ($\beta \lesssim 3^\circ$ in simulations and $\beta \lesssim 4^\circ$ in experiments) and do not allow for 
sorting in this range of lengths.

$\bullet$~In regime II ($1.3\lambda<L<2.3\lambda$ in simulations, with a slightly smaller upper bound in the experiments), the migration angle of the fibers abruptly rises to a plateau, reaching $\beta \approx 8^\circ$ in simulations and 
experiments. 

$\bullet$~When fibers become very long ($L > 2.3\lambda$ in simulations, with a slightly lower threshold in experiments), the migration angles decrease and plateau to a lower value compared to regime II, defining regime III.


Experimental and numerical results demonstrate that our tilted pillar array acts as a \textit{band-pass filter}, with strong lateral migration for intermediate fiber lengths, and weak lateral migration for short and very long fibers. 
The sorting efficiency of the pillar array is demonstrated and quantified in \textit{SI Appendix}, Fig.~S4 and Fig.~S5. 

Despite the very good general agreement between experiments and simulations, small differences remain and can be attributed to several factors.
Firstly, during transport in the array, fibers may break due to shear forces or photodamage and the fiber debris will be recovered at lateral migration distances corresponding to longer fibers. Furthermore, in experiments, fiber suspensions enter the pillar array with lateral positions that span a certain, small, range (typically $0\,\unit{\um}$ to $100\,\unit{\um}$). This variation can introduce an error of up to approximately $0.95^\circ$ in the measured value of $\beta$. Finally, imperfections in the microfluidic device, the Brownian nature of the actin filaments and the 3D deformation of fibers in the experiments also contribute to measurement uncertainties not captured by the simulations. These effects could explain the smoothing of the  transitions comparing the experimental results to numerical simulations and the large error bars observed in experiments. 

\subsection*{Migration modes and filament tension}
\label{sec:zigzagjump}
\begin{figure*}
    \centering    
    \includegraphics[width=17.8cm]{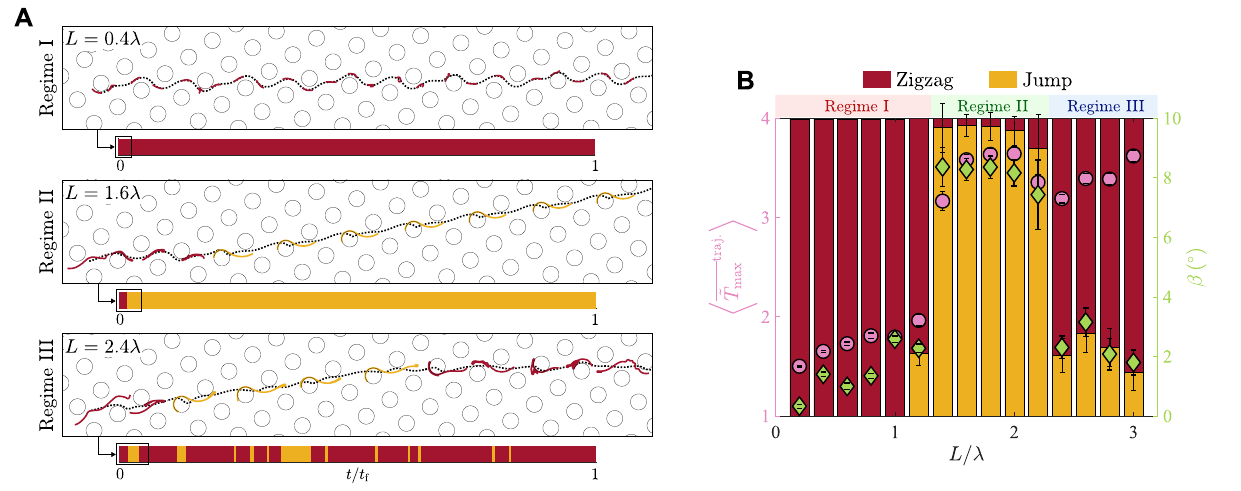}
    \caption{Chip-scale statistics. (\textit{A}) Typical examples of simulated chronophotographs and trajectories observed in the 3 regimes, along with a representation of the modes of transport of the fiber over time. The chronophotographs only show the first timesteps of the simulations. (\textit{B}) Fractions of the zigzag and jump modes averaged over 50 realizations for each fiber length. Vertical black lines show the standard deviation. Pink circular symbols represent the averaged maximum ensemble average of normalized tensions over 50 realizations for each length. Green diamond symbols denote the migration angle $\beta$. 
    }
    \label{fig:transitions}
\end{figure*}

As discussed in the previous section, cross-streamline migration enhances the lateral displacement of intermediate-length fibers. We refer to the motion of the fiber, which suddenly changes from being wrapped around one pillar to another pillar located two columns downstream, as a \textit{jump}. In contrast, all other dynamics are categorized as \textit{zigzag}. In order to connect the migration angle to the fiber dynamics in the lattice,  we show representative chronophotograph sequences from the initial time steps of each regime, with trajectories color-coded by the type of dynamics observed, in Fig.~\ref{fig:transitions}\textit{A}.  The corresponding dynamic behavior over the entire simulation is shown beneath each chronophotograph. In regime I, the fiber dynamics are entirely characterized by \textit{zigzag} motion, both in simulations and experiments (\href{run:./Video_SI/Movie S1.mp4}{\textit{SI Appendix}, Movie S1}), explaining the limited lateral displacement observed in Fig.~\ref{fig:separation_vs_length}.
The non-zero migration angle for very short fibers results from the misalignment between the \textit{zigzag} trajectories and the flow direction within the array \cite{Kim2017}.
For longer fibers in this regime, the combination of the aforementioned effect and the increased likelihood of interactions between pillars and fibers, due to their greater lengths, results in slightly higher migration angles.

In contrast, regime II is dominated by \textit{jump} events occurring after a short transient, leading to the significant lateral displacement observed in Fig.~\ref{fig:separation_vs_length}. 
The transient motion at the start depends on the fiber's initial configuration and generally accounts for less than 5\% of the simulation duration. Once the \textit{jump mode} is established, the fiber can exhibit consecutive periodic wrap-and-jump dynamics every other column  (\href{run:./Video_SI/Movie S2.mp4}{\textit{SI Appendix}, Movie S2}). The resulting migration angle (see Fig.~\ref{fig:separation_vs_length}) aligns with the geometric inclination of the two pillars relative to the flow direction: $\beta\approx\alpha-\arctan(\lambda/2\lambda) = 35^\circ - \arctan(0.5)\approx 8.4^\circ$ (with the $\arctan$ in degree).

In regime III, both \textit{zigzag} and \textit{jump} motions occur alternately, with \textit{zigzag} being more frequent overall (\href{run:./Video_SI/Movie S3.mp4}{\textit{SI Appendix}, Movie S3}). This implies that very long fibers in regime III experience relatively small lateral displacement. Notably, these fibers exhibit complex configurations and undergo dramatic deformations, such as folding, during their \textit{zigzag} motion.
Note that, in the simulations, the transition between regimes II and III occurs near $L \approx 2.3 \lambda$ (see Fig.~\ref{fig:separation_vs_length}\textit{B}), which is very close to the distance between two consecutive pillars in the jump mode  $\sqrt{5}\lambda$. This close agreement indicates that these complex, chaotic, deformations are triggered by interactions with multiple pillars simultaneously.

Fig.~\ref{fig:transitions}\textit{B} shows the fractions of \textit{zigzag} and \textit{jump} motions as a function of fiber length, displaying the same trend as the band-pass effect observed in Fig.~\ref{fig:separation_vs_length}. 
Furthermore, we average the maximal normalized tension in the fiber, as shown in Fig.~\ref{fig:tension_during_wrapping_phase}\textit{B}, for each trajectory and calculate the mean of this temporal average for the 50 simulations for each length. This tension (shown by the pink full circles in Fig.~\ref{fig:transitions}\textit{B}) shows a strong increase at the transition between regime I and regime II, further confirming, from a statistical perspective, that tension plays a crucial role in fiber cross-streamline migration. The transition towards regime III is less marked for the tension since fiber dynamics becomes increasingly complex in regime III and tension is not the only factor that drives lateral displacement. However, the slight drop in the average tension between regime II and III might indicate the decreasing importance of fiber pillar contact and the change in transport dynamics.

\section*{Discussion}
\label{sec:discussion}

To summarize, by combining experiments and numerical simulations, we demonstrated the potential of using a regular array of circular pillars arranged in a square lattice for separating  flexible fibers based on their length. In contrast to traditional DLD techniques for particle separation, we observed that our device is a band-pass filter isolating fibers of intermediate length.

In our study, we varied the tilt angle $\alpha$ of the lattice and analyzed its impact on the lateral migration of the fibers. When $\alpha = 0^\circ$ or $45^\circ$, the flow maintains symmetry, and no separation of fibers is observed. For other tilt angles, the symmetry of the flow is broken, leading to the separation of short and long fibers. We found that fiber separation is maximized at a tilt angle of $\alpha = 35^\circ$, both in the experiments and simulations. We rationalized this ``optimal" angle by the fact that at this value of the tilt angle, streamlines are separated in groups of similar spatial extension. This angle slightly differs from the earlier numerical investigation of flexible fiber sorting in pillar arrays, which focused on chromatographic sorting and concluded that $\alpha=30^\circ$ is the optimal angle based on the average separation distance, used for chromatography, rather than  the lateral drift commonly used for DLD \cite{chakrabarti2020trapping}.

At this optimal tilt angle, the microfluidic device operates as a band-pass filter with a bandwidth of approximately $0.7 \lambda$. The separation arises from the fibers’ length-dependent preference for either the \textit{zigzag} or \textit{jump} mode of motion. Fibers shorter than $1.3\lambda$ predominantly follow the zigzag mode, exhibiting small lateral migration with a drift angle $\beta \lesssim 3^\circ$ in simulations and $\beta \lesssim 4^\circ$ in experiments. 
Fibers with lengths in the range $1.3\lambda < L \lesssim  2\lambda$ are sufficiently long to change groups of streamlines after the wrapping phase around a pillar. They follow the jump mode, jumping from pillar to pillar with a trajectory angle $\beta\approx 8^\circ$, given by the geometry of the pillar array.
Longer fibers ($L \gtrsim 2\lambda$) switch intermittently between the zigzag mode and the jump mode, with a net preference for the zigzag mode. Similar to shorter fibers, they have a small migration angle ($\beta \approx 2^\circ$ to $3^\circ$) in simulations. However, in experiments, the increasing complexity of fiber dynamics with greater length results in a less distinct boundary between the two regimes, and the migration angle for longer fibers can be $\beta \gtrsim 3^\circ$. 

Furthermore, numerical simulations revealed that internal tension, generated through viscous forces pulling on the fiber when it is wrapped around and delayed by the pillars, is the key mechanism driving cross-streamline migration. This tension effectively rigidifies the fiber and hinders its deformation along the local streamlines, ultimately leading to a net lateral displacement. Statistically, intermediate-length fibers exhibit the highest internal tension, correlating with the strongest lateral migration. Despite the fact that fiber dynamics remains complex and dependent on multiple factors, we believe this behavior to be at the origin of the observed band-pass effect. 

Using these findings, a highly efficient separation device can be envisioned, combining two pillar arrays with varying gap sizes, and a fixed tilt angle of $\alpha = 35^\circ$, in series in a microchannel, as shown in \textit{SI Appendix}, Fig. S3. This design enables the precise collection of fibers within a desired length range by leveraging the intersection of the band-pass effects from both arrays.
 In addition, our results could also be exploited to improve selective sorting of flexible polymers, such as DNA, in DLD devices under flow, by including their transient flow-induced elongated shapes. Future investigations could focus on optimizing the lattice layout and pillar geometries, along with conducting a more comprehensive analysis of fiber deformability, to further enhance the device's versatility and separation efficiency. This study highlights the potential of DLD for fiber sorting and establishes a foundation for advancing microfluidic separation technologies.








\matmethods{

\subsection*{Materials and experimental methods}
We perform experiments in polydimethylsiloxane (PDMS) microchannels with a pillar array that spans the entire channel width ($W_{\rm array} = W_{\rm ch}=400\,\unit{\um}$) and  extends approximately $L_{\rm array}=5\,\unit{\mm}$ along the channel, positioned in the center, 
as shown in Fig.~\ref{fig:setup}\textit{A}. The arrays are arranged in a square-lattice layout with a lattice size of $\lambda = 30\,\unit{\um}$, oriented at angles (tilt angles) ranging from $\alpha = 0^{\circ}$ to $\alpha = 45^{\circ}$ relative to the channel length (main flow direction). All the pillars extend throughout the entire depth of the channel, $H_{\rm ch}=50\,\unit{\um}$, and the radius of the pillar is $R = 10 \pm 1\,\unit{\um}$, as shown in the magnified top view in Fig.~\ref{fig:setup}\textit{A}.

We use actin filament as the experimental fiber. The synthesis of actin filaments is based on a well-controlled and reproducible protocol detailed extensively in \cite{Liu2018MorphologicalFlow}. The filaments in our
experiments are labeled fluorescently and stabilized with phalloidin. They have typical lengths of 5 to $100\,\unit{\um}$, diameters of 5 to $9\,\unit{\nm}$ and a persistence length of $\ell_{\rm p} = 17 \pm 1\,\unit{\um}$, as measured by analysis of thermal-fluctuation-induced conformational changes \cite{Liu2018DynamicsFlow}. $45.5\,\unit{\percent}$ (w/v) sucrose is added to the solution to match the refractive index of the PDMS channel, and the viscosity of the suspension is around $5.6\,\unit{\milli\pascal\second}$ at 20\,\unit{\degreeCelsius}.

To maintain a consistent flow during the experiments, we inject the filament suspension at the channel's inlet, at $Q = 1\,\unit{\nano\litre}$/s, while simultaneously using a syringe at the outlet to extract liquid at the same flow rate.
Actin filament images are captured using a high-performance 63\texttimes\, water immersion objective with a large numerical aperture (Objective C-Apochromat 63\texttimes/1.2 W autocorr M27, Zeiss), resulting in a field of view of approximately \numproduct{200x200} \unit{\um}. 
Our measurement plane is positioned at the mid-height of the channel to minimize out-of-plane shear.
In the first part of our pore scale study, we focus on the local flow and fiber dynamics within and around a unit cell of the arrays, positioning the region of interest at the center of the pillar array to minimize wall effects, as illustrated in Fig.~\ref{fig:setup}\textit{A}.

To investigate fiber suspension separation and the band-pass effect under the optimal tilt angle, we design an additional chip-scale statistical experiment (see \textit{SI Appendix}, Fig. S1) with a tilt angle of $\alpha = 35^{\circ}$.

\subsection*{Numerical modeling}
The simulations are performed in three dimensions in a unit cell of height $H_{\rm ch}$ consisting of 4 quarter circular cylinders of radius $R$ arranged in a square layout, with center-to-center distance $\lambda$. They are confined between two parallel walls positioned in the $xy$-plane at $z = \pm H_{\rm ch}/2$.

A flexible fiber of length $L$ and radius $a$ is immersed in the midplane of the unit cell (dotted square in the left panel of Fig.~\ref{fig:setup}\textit{A}).
It is transported by a fluid flow of imposed incident direction $-\alpha$ and maximum velocity $U_{\rm max}$.
As the Reynolds number reported in the experiments is small, the flow is governed by the Stokes equation
\begin{equation}
\nabla p - \rho{\bf f} = \mu \nabla ^2 {\bf u}
\label{eq:Stokes_equation}
\end{equation}
where $p$ is the pressure, $\rho$ the density, $\mu$ the dynamic viscosity, ${\bf u}$ the velocity of the fluid, and ${\bf f}=(f_0\cos\alpha,-f_0\sin\alpha,0)$ is the body force that plays the role of a pressure gradient and triggers the flow. 
No-slip boundary conditions are set on the channel  walls and on the surface of the pillars.
The unit cell is open in the $y$ and $x$ directions where periodic boundary conditions are applied for the geometry, the flow and the fiber to model the periodic pillar array.

The approach we use to compute the flow and the transport of the fiber is explained in details in our previous work \cite{Li2024}, we thus here only briefly introduce the key ingredients of the numerical method.

The flow field inside the unit cell is computed in three dimensions using the lattice Boltzmann method (LBM) \cite{Krueger2016,Succi2001} with a D3Q19 lattice and the Bhatnagar-Gross-Krook collision operator \cite{Bhatnagar1954}.
This flow transports the fiber within the pillar array and exerts mechanical stress on it.
As a reaction to this stress, the fiber experiences internal stretching and bending forces.
The fiber is also subjected to lubrication and steric forces when it approaches the pillars closely.
The interplay between all these forces determines the dynamics and the trajectory of the fiber.

In the simulations, we use a bead-spring model to account for these elastohydrodynamic couplings.
This bead-spring model has been extensively used and validated in the literature to study the motion of rigid and flexible fibers in viscous flows \cite{yamamoto1993method,schlagberger2005orientation,wada2009hydrodynamics,manghi2006propulsion,delmotte2015general,marchetti2018deformation,schoeller2021methods,slowicka2022buckling,Makanga2023,Li2024}.

The fiber is modeled by a set of $n$ rigid beads that are connected together by internal elastic forces, ${\bf F}^{\rm E}$, derived from an elastic potential $H$ \cite{Gauger2006,marchetti2018deformation,Makanga2023}
\begin{align}
    {\bf F}^{\rm E} = -\nabla H
\end{align}
in which
\begin{align}
    &
    H = \sum_{i=2}^n \left[ \frac{S}{4a} \left( \left|{\bf t}_i\right| - 2a\right)^2 \right] + \sum_{i=2}^{n-1}\left[ \frac{B}{2a}\left(1 - {\bf \hat t}_{i+1} \cdot {\bf \hat t}_i \right) \right]
    &
\end{align}
where the first sum accounts for stretching forces and the second one for bending forces.
The mechanical response of the fiber to stress is dictated by the stretching coefficient $S = E \pi a^2$ and the bending coefficient $B = E \pi a^4 /4$, with $E$ the Young's modulus of the fiber.
${\bf t}_i$ is the vector connecting the center of mass (CoM) of the $i$th and $(i-1)$th beads, and ${\bf \hat t}_i = {\bf t}_i/\left|{\bf t}_i\right|$ is the unit tangent vector between these two beads.
The dimensionless internal tension along the fiber centerline, $\tilde{T}$, is given by the stretching force acting on each spring in between pairs of beads normalized by the reference drag force $F_{\rm ref}=6\pi\mu a U_{\rm max}$.

When the fiber approaches the surface of a pillar closely, it experiences lubrication forces that tend to slow down its motion.
The normal and tangential lubrication forces, ${\bf F}^{\rm L}_{{\rm n}}$ and ${\bf F}^{\rm L}_{{\rm t}}$, are respectively based on the analytical solutions provided by Cooley and O'Neill \cite{Cooley1969}, and Goldman, Cox and Brenner \cite{Goldman1967} for spheres approaching walls in viscous flows.
They only apply when the gap $\delta$ between a fiber bead and the surface of a pillar is smaller than the bead radius, i.e. when $\delta/a < 1$.
The normal and tangential lubrication forces acting on bead $i$ due to pillar $j$ are computed as
\begin{align}
{\bf F}^{\rm L}_{{\rm n},ij} = 6\pi\mu\frac{a^2}{d}U_i^\perp{\bf n} & & {\bf F}^{\rm L}_{{\rm t},ij} = \frac{16}{5}\pi\mu a \ln{\frac{d}{a}}U_i^\parallel {\bf t}
\end{align}
with $U_i^\perp$ and $U_i^\parallel$ the normal and tangential components of the velocity of bead $i$, and ${\bf n}$ and ${\bf t}$ the unit vectors normal and tangent to the pillar surface.
To avoid numerical instabilities when the gap $\delta$ between the fiber bead and the surface of the pillar becomes too small, we define $d=\max\left(\delta,d_{\rm min}\right)$, with $d_{\rm min}=0.8a$.
This value of $d_{\rm min}$ is quite high and makes the lubrication force reach a plateau rapidly, but it allows us to model an effective lubrication force that strongly slows down the fiber when it approaches a pillar, as observed in the experiments.
The lubrication forces also contribute to avoid artificial overlapping between the fiber and the pillars.
In addition to lubrication forces, a repulsive force is added when the fiber approaches a pillar closer than a given cutoff $R_{\rm ref}$ to ensure it does not penetrate inside the pillars.
The repulsive force ${\bf F}_{ij}^{\rm R}$ acting between fiber bead $i$ and pillar $j$ is based on the one used in \cite{Dance2004CollisionBE} and computed as
\begin{align}
{\bf F}^{\rm R}_{ij} = \begin{cases}
    -\frac{F_{\rm ref}}{a}\left[ \frac{R^2_{\rm ref} - \left|{\bf r}_{ij}\right|^2}{R^2_{\rm ref} - a^2} \right]^{2}{\bf r}_{ij} & \text{if}~ \left| {\bf r}_{ij} \right| < R_{\rm ref} \\
    {\bf 0} & \text{otherwise}
  \end{cases}
  \label{eq:repulsive_force}
\end{align}
where ${\bf r}_{ij}$ is the center-to-center distance between bead $i$ and pillar $j$.

In the simulations we set $R_{\rm ref} = R + 1.1a$. This cutoff distance is small enough to allow the fiber to get close to the no-slip condition on the pillars surface, and large enough to prevent artificial overlapping.
A repulsive force similar to \eqref{eq:repulsive_force} (with $a$ replaced by $2a$) is applied to fiber beads that are separated by less than $2.2a$ to prevent the fiber from overlapping itself.
The total repulsive force acting on the $i$th fiber bead is then the sum of the repulsive forces over the four quarter pillars of the unit cell and over all the fiber beads.
Since it is difficult to quantify or control the exact contact conditions in experiments, solving for the exact flow in the thin gap between the fiber and obstacle surfaces is impossible due to their
unknown topography and tribology. Our simple model, which uses lubrication and a repulsive force with a cutoff distance, offers a robust alternative that maintains numerical stability and produces results that are in excellent agreement with experiments.


Once the internal and external forces are computed, the velocity of the fiber beads is obtained from the mobility relation \cite{guazzelli2011physical}
\begin{equation}
    {\bf U}_i = {\bf u}_i +  \sum_{j=1}^n {\bf M}_{ij} {\bf F}_j 
    \label{eq:bead_vel}
\end{equation}
where ${\bf U}_i$ is the velocity of bead $i$, ${\bf u}_i$ is the bead velocity induced by the background flow, ${\bf M}$ is the mobility matrix accounting for all hydrodynamic interactions (HI) between the fiber beads, and ${\bf F}_j$ the non-hydrodynamic forces acting on bead $j$.
In this work, we use the Rotne-Prager-Yamakawa mobility matrix \cite{Wajnryb2013}, and the velocity ${\bf u}_i$ is evaluated by interpolation of the background flow at the CoM of bead $i$ using a kernel that approximates well the no-slip condition for a bead of radius $a=\Delta x$, with $\Delta x$ the spacing of the LBM grid. 
The mobility matrix ${\bf M}$ accounts for the disturbances induced by the fiber on the flow, and for its drag anisotropy, through HI between the fiber beads; but it neglects the corrections of these HI due to the channel walls and obstacles' surface. However, as shown by the excellent agreement with experiments, and as discussed in Appendix A of our previous work \cite{Li2024}, these corrections are small compared to the fiber velocity induced by the ambient flow and to the lubrication corrections in the near field.
}

\showmatmethods{} 

\acknow{AL and OdR acknowledge funding from the ERC Consolidator Grant PaDyFlow (Agreement  682367). This work has received the support of the Institut Pierre-Gilles de Gennes (Investissements d’avenir ANR-10-EQPX-34) and
 IdEx (ANR-18-IDEX-0001). ZL acknowledges funding from the Chinese Scholarship Council. BD and CB acknowledge support from the French National Research Agency (ANR), under award ANR-20-CE30-0006. OdR acknowledges funding from the French National Research Agency (ANR) under award ANR-21-CE13-0048 and ANR-22-CE30-0024. OdR is member of GDR 2108 Quantitative approach of Life.}

\showacknow{} 

\bibsplit[21]

\bibliography{fiber_bandpass}

\end{document}